\documentclass[a4paper,fleqn,usenatbib]{mnras}

\usepackage[T1]{fontenc}
\usepackage{ae,aecompl}

\usepackage{graphicx}	% Including figure files
\usepackage{amsmath}	% Advanced maths commands
\usepackage{amssymb}	% Extra maths symbols
\usepackage{color}
\usepackage{colortbl}
\title[Radioactive elements]{Chemical Evolution of $^{26}$Al and $^{60}$Fe in the Milky Way}

\author[Vasini et al.]{A. Vasini$^{1,2}$\thanks{E-mail:
		arianna.vasini@inaf.it}, F. Matteucci$^{1, 2,3}$, E. Spitoni$^{4,5,6}$\\
	$^1$ Dipartimento di Fisica, Sezione di Astronomia, Universit\`a di
	Trieste, via G.B. Tiepolo 11, I-34131, Trieste, Italy \\
	$^2$ I.N.A.F. Osservatorio
	Astronomico di Trieste, via G.B. Tiepolo 11, I-34131, Trieste,
	Italy\\
	$^3$ I.N.F.N. Sezione di Trieste, via Valerio 2, 34134 Trieste, Italy\\
	$^4$ Universit\`e C$\hat{o}$te d'Azur, Observatoire de la C$\hat{o}$te
	d'Azur, CNRS, Laboratoire Lagrange, Bd de l'Observatoire, CS 34229,\\06304
	Nice Cedex 4, France \\
	$^5$Konkoly Observatory, Research Centre for Astronomy and Earth Sciences, Konkoly Thege Mikl\'{o}s \'{u}t 15-17,\\ H-1121 Budapest, Hungary\\
	$^6$ Stellar Astrophysics Centre, Department of Physics and Astronomy, Aarhus University, Ny Munkegade 120,\\ DK-8000 Aarhus C, Denmark\\
}

\begin{document}
	\date{Accepted . ; in original form xxxx}
	
	\pagerange{\pageref{firstpage}--\pageref{lastpage}} \pubyear{xxxx}
	
	\maketitle
	
	\label{firstpage}
	
	\begin{abstract}
		
		We present theoretical mass estimates of $^{26}$Al and $^{60}$Fe throughout the Galaxy, performed with a numerical chemical evolution model including detailed nucleosynthesis prescriptions for stable and radioactive nuclides. We compared the results for several sets of stellar yields taken from the literature, for massive, low and intermediate mass stars, nova systems (only for $^{26}$Al) and supernovae Type Ia.We then computed the total masses of $^{26}$Al and $^{60}$Fe in the Galaxy. We studied the bulge and the disc of the Galaxy in a galactocentric radius range 0-22 kpc. We assumed that the bulge region (within 2 kpc) evolved quickly suffering a strong star formation burst, while the disc formed more slowly and inside-out.
		We compared our results with the $^{26}$Al mass observed by the $\gamma$-ray surveys COMPTEL and INTEGRAL to select the best model. Concerning $^{60}$Fe, we do not have any observed mass value so we just performed a theoretical prediction for future observations.
		In conclusion, low, intermediate mass stars and Type Ia supernovae contribute negligibly to the two isotopes, while massive stars are the dominant source. The nova contribution is, however, necessary to reproduce the observations of $^{26}$Al. Our best model predicts $2.12$ M$_{\odot}$ of $^{26}$Al, in agreement with observations, while for $^{60}$Fe our best mass estimate is $\sim 1.05$ M$_{\odot}$.
		We also predicted the present injection rate of $^{26}$Al and $^{60}$Fe in the Galaxy and compared it with previous results, and we found a larger present time injection rate along the disc.
		
	\end{abstract}
	
	
	\section{Introduction}
	
	The $\gamma$-ray Astronomy maps the emission in the $\gamma$ band produced throughout the galaxies. The latest $\gamma$-surveys, COMPton TELescope (COMPTEL, see Sch\"{o}nfelder et al. $1984$) and INTErnational Gamma-Ray Astrophysics Laboratory (INTEGRAL, see Winkler $1994$), were both designed to detect the decay emission in the Galaxy of two unstable elements, $^{26}$Al and $^{60}$Fe. They are both short-lived radioisotopes (from now on SLRs), with decay time $\tau_{26Al}=1.05$ Myr and $\tau_{60Fe}=3.75$ Myr for $^{26}$Al and $^{60}$Fe, respectively (Diehl $2013$). Their lifetimes make them suitable for the study of the recent history of the Milky Way, with particular focus on the latest nucleosynthesis events (see Brinkman et al. $2021$). These elements are considered tracers of the regions of active star formation (Limongi \& Chieffi $2006$): massive stars are the main contributors to their Galactic abundance, so the present observations of $^{26}$Al and $^{60}$Fe mark the regions where stars were formed within the last millions of years. For the same reason, they are also tracers of the ejects of Supernovae (SNe) core-collapse (Type II, Ib, Ic): the decay emission lines help in highlighting the flow of material through the ISM after the explosion. 
	\newline
	Between $1991$ and $2000$ the imaging telescope COMPTEL performed an all-sky survey to detect the $1.809$ MeV emission line produced during the $^{26}$Al decay into $^{26}$Mg (Diehl $1995$). The detection system was composed of two detector planes: at first, a photon Compton scattered on the upper plane and then it was absorbed by the lower plane. The coincidence of the two events determined the detection of a $1.809$ MeV photon. The data analysis highlighted the presence of a diffuse emission at $1.809$ MeV on the Galactic plane, with some emission spots superimposed. Quantitatively, COMPTEL estimated around $1.5-2$ M$_{\odot}$ of $^{26}$Al within $5$ kpc from the Galactic centre (Prantzos \& Diehl, $1996$). 
	\newline
	Later, in $2002$ INTEGRAL performed measurements both for $^{26}$Al and $^{60}$Fe (Diehl $2013$). The data collected for $^{26}$Al confirmed the earlier COMPTEL results: the emission is diffused and the $^{26}$Al mass within $5$ kpc from the Galactic centre ranges from $1.7\pm 0.2$ M$_{\odot}$ (Martin et al. $2009$) to $2.0\pm 0.3$ M$_{\odot}$ (Diehl et al. $2010$, Diehl $2016$), depending on the assumed source location. Concerning $^{60}$Fe, only its flux was measured: the brightness of its two emission lines, $1.173$ MeV and $1.332$ MeV, was $\sim 15\%\pm 5\%$ of that of $^{26}$Al. The irradiation by cosmic rays in the spacecraft produced $^{60}$Co nuclei which interfered with the instruments, preventing a $^{60}$Fe mass estimate.
	\newline
	Unfortunately, the progresses in $\gamma$-ray observations were not followed at the same rate by chemical evolution investigations, which could have offered new theoretical constraints independent of the observational ones. One of the main reasons for this lack is the complexity of the treatment of radioactive nuclei within a standard chemical evolution model. To obtain easy and explicit solutions for the equations, the adopted models were mainly analytical, such as those developed by Clayton ($1984$) and Clayton ($1988$). The limits of analytical models are the assumptions necessary to derive the solution which are typically very restrictive. For example, some models consider only one burst of star formation, simulating a stellar population formed at the same time and from the same gas, a very rough hypothesis for the evolution of the Milky Way. Another restrictive hypothesis is IRA (instantaneous recycling approximation): stars with  mass  $<1$ M$_{\odot}$ never die, whereas stars $>1$ M$_{\odot}$ die instantaneously, thus enriching immediately the interstellar medium (ISM) with their nucleosynthesis products. In other words, this approximation neglects the stellar lifetimes and is acceptable only for elements produced on timescales negligible relative to the age of the Universe.
	
	The most  robust way to study chemical evolution is by using numerical models, both for stable and unstable nuclei: in these models we allow for many stellar generations and different histories of star formation and we take into account in detail the stellar lifetimes and stellar yields.
	In the past, detailed numerical models for radioactive nuclides have been computed by Timmes et al. (1995) ($^{26}$Al and $^{60}$Fe) and by C\^{o}t\'{e} et al. ($2019$), this latter  focused on the isotopic ratios at the time of formation of the Solar System.

	%What can be done for radioisotopes is adding them to an already existing stable element model. In this way the stable elements work as constraints, ensuring more reliable results for the unstable ones. 
	This work is aimed at analyzing the chemical evolution of $^{26}$Al and $^{60}$Fe in the Milky Way by means of a very detailed chemical evolution model, already tested for a wide set of data relative to both the solar vicinity and the whole disc (see Romano et al. 2010; 2020, Palla et al. 2020). The model, accounting for the formation and evolution of the thick- and thin- disc, is a revised version of the two-infall model of Chiappini et al. ($1997$). For the bulge we adopt a separate model taken from Matteucci et al. ($2019$). Both models (for discs and bulge) take into account detailed stellar lifetimes, SN progenitors and detailed stellar yields. Moreover, they assume that the two discs and bulge formed by different episodes of extragalactic gas accretion, occurring on different timescales. Here, these models have been improved by including the equations for the radioactive elements $^{26}$Al and $^{60}$Fe. The adopted stellar yields are the most recent ones and they include: massive stars, Asymptotic Giant Branch (AGB) stars, Super-AGB stars, Type Ia SNe and, for the first time, the contribution of novae to radioactive nuclide production.
	We aim at obtaining new theoretical constraints to the observed mass of $^{26}$Al in the Galaxy, as well as providing a clear prediction for the $^{60}$Fe mass in the Milky Way that could be used as a reference for future $\gamma$ observations. We also intend to compare results with the previous work on $^{26}$Al and $^{60}$Fe by Timmes et al. (1995). 
	\newline
	The paper is structured as follows: in Section $2$ we describe the chemical evolution model, in Section $3$ we show the results obtained, in Section $4$ we compare some theoretical predictions with the observational constraints and in Section $5$ we draw the main conclusions.
	
	\section{The model}
	In this Section we present the main assumptions and characteristics of the adopted chemical evolution model for the different galactic components (discs and bulge).
	\subsection{Main assumptions}
	\subsubsection{Thick and thin discs}
	We aim at studying the evolution of the chemical abundances of $^{26}$Al and $^{60}$Fe throughout the whole Galaxy, both in the bulge and in the discs.
	To purse this objective, for the thick- and thin-disc evolution we consider the two-infall model as described by Chiappini et al. ($1997$) and Chiappini et al. ($2001$), then revised by Romano \& Matteucci ($2003$) and Romano et al. ($2010$; $2020$) and more recently by Grisoni et al. ($2017$) and Spitoni et al. ($2019$).
	During the first fast infall episode (lasting no longer than 1 Gyr) the thick disc forms, followed by the second accretion event which forms the thin- disc and occurs on much longer timescales which increase with Galactocentric distance (inside-out formation, see Matteucci \& Fran\c{c}ois, $1989$). The Galactic thin-disc is divided into concentric rings around the Galactic centre, each of them $2$ kpc wide, without any exchange of matter among them. The first ring inside $2$ kpc contains the bulge. Inside each ring, homogeneous mixing of gas is assumed. The total surface mass density in each ring is tuned to reproduce the observed exponential present time distribution (see later). The amount of gas and its chemical composition are instead the unknowns of our model.
	\newline
	To describe the Galactic thin-disc ($R> 2$ kpc) we assumed the star formation rate (SFR) as suggested by Schmidt-Kennicutt, based on observational data of local star forming galaxies (Kennicutt $1998$) :
	\begin{equation}
	SFR(r,t)\propto\nu\sigma_{gas}^{k}(r,t)
	\end{equation}
	where $\sigma_{gas}(r,t)$ is the gas surface mass density and $k=1.4$ the law index. The parameter $\nu$ is the star formation efficiency, namely the SFR per unit mass of gas, which is set to $\nu= 1$ Gyr$^{-1}$. This value of the efficiency of star formation ensures that the present time SFR is compatible with the observed values. For the thick-disc we assume the same law but with an efficiency of star formation $\nu=2$ Gyr$^{-1}$, in agreement with previous works (e.g. Grisoni et al., $2017$).
	\newline
	The IMF adopted is the three-slopes power law by Kroupa et al. ($1993$) for the solar vicinity:
	\begin{equation}
	IMF_{K}(M)\\\propto\left \{ \begin{array}{rl}
	M^{-0.3}\,\,\,\,\,\,\,\,\,\,\,\,\,\,\,\,\,\,\,\,\,\,\,M<0.5 M_{\odot}\\
	M^{-1.2}\,\,\,\,\,\,\,\,\,\,0.5<M/M_{\odot}<1\\
	M^{-1.7}\,\,\,\,\,\,\,\,\,\,\,\,\,\,\,\,\,\,\,\,\,\,\,\,\,\,\,\,M>1 M_{\odot}
	\end{array}
	\right.
	\end{equation}
	The IMF is assumed to be valid in the mass range $0.1-100$ M$_{\odot}$, as in the majority of chemical evolution models.
	\newline
	The two-infall law is assumed to be (see Matteucci $2021$):
	\begin{equation}
	A(R,t)=a(R)e^{-t/\tau_{T}}+b(R)e^{-(t-t_{max})/\tau_{D}(R)}	
	\end{equation}
	The parameters $a(R)$ and $b(R)$ are tuned to reproduce the present time total surface mass density of the thick  and thin disc, respectively. We assumed as total present day surface mass density in the solar vicinity $\sigma$= 54 M$_{\odot}$ pc$^{-2}$ (Romano et al. $2000$). The parameters $\tau_{T}$  and $\tau_{D}(R)$ are the timescales of the two infall episodes, expressed in Gyr, where the first is related to the formation of the thick-disc, while the  second one is connected with the thin-disc and varies with Galactocentric radius. We assumed, in agreement with previous works  $\tau_{T}=1$ Gyr and we computed $\tau_{D}(R)$ using the relation $\tau_{D}(R)=1.033\cdot(R/kpc)-1.267$ Gyr to account for the \textquotedblleft inside-out\textquotedblright $\,\,$scenario (Chiappini et al. $2001$). Moreover, $t_{max}=1$ Gyr is the time of the maximum infall in the second accretion episode: in other words, it indicates the delay between the end of the first and the beginning of the second infall episode.
	In Figure $1$ we show the SFR as a function of time in the solar neighborhood. The two peaks are due to the assumed double infall. The first infall, with $1$ Gyr of time scale, during which the thick-disc formed, is responsible for the first peak, whereas the second infall, still ongoing at the present time formed the thin-disc. The present SFR value at $8$ kpc is around $\sim 3$ M$_{\odot}$ pc$^{-2}$ Gyr$^{-1}$, in agreement with the observations (see Prantzos et al. $2018$).
	\begin{figure}
		\centering
		\includegraphics[scale=0.5]{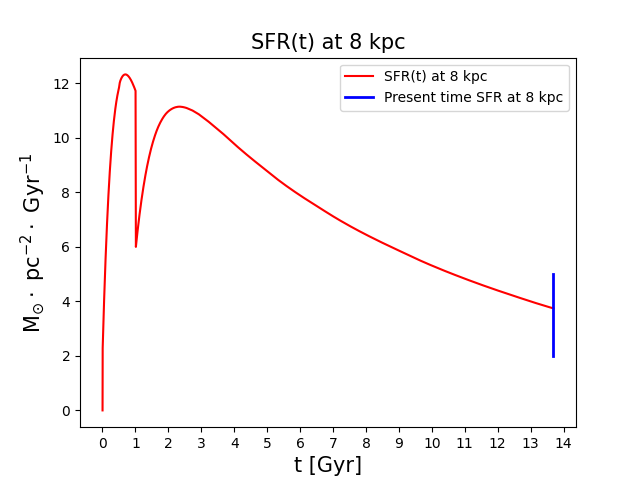}
		\caption{The red line represents the SFR(t) at $8$ kpc as a function of time. The SFR in the solar neighborhood, according to the prescriptions assumed, shows two peaks: the first infall is responsible of the first SFR peak, then SFR drops until the second infall starts and SFR increases again. The blue line marks the observed present time SFR according to Prantzos et al. ($2018$).}
		\label{fig:sfr8kpc}
	\end{figure}
	\newline
	Since SNe Ia and core-collapse, as well as novae, are considered as producers of the elements $^{26}$Al and $^{60}$Fe, we describe here how we do compute their rates.
	
	The SNIa rate is given by (see Matteucci \& Greggio, $1986$; Matteucci \& Recchi, $2001$):
	\begin{equation}
	R_{SNIa}(t)=A_{B}\int_{B_{Bm}}^{M_{BM}}\varphi(m)\bigg[\int_{\mu_{B_{min}}}^{0.5}f(\mu_{B})\psi(t-\tau_{m2})\bigg] dm	
	\end{equation}
	where $A_{B}=0.09$ is the fraction of binary systems able to produce a SNIa assuming a single degenerate scenario, and is chosen to reproduce the present time Type Ia SN rate. The mass $M_{Bm}$ is a free parameter, $M_{BM}=16$ M$_{\odot}$, $f(\mu_{B})$ is the distribution of the mass ratio $\mu_{B}=M_{2}/M_{B}$ between the two companions and $\tau_{m2}$ is the lifetime of the secondary star, that dies later and represents the clock of the binary system. The term $Q_{mi}(t-\tau_{m2})$ is the production matrix as defined by Talbot \& Arnett ($1973$): it is the fraction of element $i$ ejected by a star of mass $m$, both newly formed by the star or already present in the gas out of which the star was formed.
	\newline
	The core-collapse SN rates are expressed as:
	\begin{equation}
	R_{SNII,Ib,Ic}(t)=\int_{M_{L}}^{M_{U}}\psi(t-\tau_{m})\varphi(m)dm
	\end{equation}
	with $M_{L}=8$ M$_{\odot}$, $M_{U}=30$ M$_{\odot}$ for SNe II, and$M_L$= $30$ M$_{\odot}$  and  $M_{U}=100$ M$_{\odot}$ for SNe Ib and Ic, and $\tau_{m}$ represents the stellar lifetime, adopted from Romano et al. ($2010$).
	Concerning the rate of novae (binary systems with a white dwarf plus a low mass companion), we assumed
	the formation rate of nova systems as described by Romano \& Matteucci ($2003$).
	In particular, the rate of nova formation at a given time $t$ is computed as the fraction $\alpha$ of the rate of formation of white dwarfs:
	\begin{equation}
	R_{novae}(t)=\alpha\int_{0.8}^{8}\psi(t-\tau_{m}-\Delta t)\varphi(m)dm
	\end{equation}
	as it was originally suggested by D'Antona \& Matteucci ($1991$). The parameter $\alpha$ indicates the fraction of binary systems which can give rise to nova systems and is chosen to be $\alpha=0.024$. With this value we obtain a present time nova rate of $43$ nova yr$^{-1}$ (see Table $5$), in agreement with observations in the Galaxy (Della Valle \& Izzo, 2020, and references therein; Shafter, $2017$). The time $\Delta t$ takes into account the fact that the white dwarfs in nova systems should be cool, and it corresponds to $1$ Gyr.
	
	The rate of nova outburts is then obtained by multiplying the rate of nova system formation for the estimated number of outbursts during the life of a nova system:
	\begin{equation}
	R_{novaout} (t)=10^{4} \cdot R_{novae}(t)
	\end{equation}
	
	where the number of outbursts ($10^{4}$) is taken by (Ford, $1978$).
	
	\subsubsection{The bulge}
	As already stated, the previous assumptions are valid only for the Galactic thick- and thin-disc ($R> 2$ kpc), whereas for the bulge ($R\le 2$ kpc) the prescriptions are different. We follow the model by Matteucci et al. ($2019$) , which already reproduces the main observational bulge features.
	\newline
	The SFR is again that by Kennicutt ($1998$) as reported in equation ($1$), but with a higher star formation efficiency
	$\nu=25$ Gyr$^{-1}$ and the IMF by Salpeter ($1955$), which is more top-heavy than the one for the discs and is in agreement with the IMF derived for the bulge by Calamida et al. ($2015$).
	\newline
	In particular:
	\begin{equation}
	IMF_{S}(M)\propto M^{-2.35}
	\end{equation}
	valid in the mass range $0.1-100$ M$_{\odot}$.
	These choices are required if one wants to reproduce the stellar metallicity distribution function as well as the abundance patterns of the bulge (see Cescutti \& Matteucci ($2011$), Matteucci et al. ($2019$)). 
	
	We assumed that the bulge formed very quickly during the first infall event which involved the thick-disc, so the accretion law is:
	\begin{equation}
	A(R,t)=a(R)e^{-t/\tau_{B}}
	\end{equation}
	where $\tau_{B}=0.1$ Gyr, shorter than for the thick and thin discs.
	\newline
	This choice of the input parameters outlines a bulge with an intense star formation at the beginning of its history, which leads to a quick consumption of gas with the consequence of almost no star formation at the present time. The SFR in the bulge during the first $3$ Gyr is shown in Figure $2$, where one can see that the gas in the bulge is very quickly consumed and star formation becomes very low already before 1 Gyr since the beginning of star formation, and it maintains the same value of $\sim 0.2$ M$_{\odot}$ pc$^{-2}$ Gyr$^{-1}$ up to the present time, in agreement with observations. In fact, the observed number of stars with ages lower than 5 Gyr should be not higher than $\sim 10\%$ of the total (Bernard et al. $2018$) and the youngest ones could have been accreted from the inner thin- disc (see Matteucci et al. $2019$). Therefore, the contribution of the bulge to the present time observed masses of $^{26}$Al and $^{60}$Fe, which have decay timescales of the order of few million years, is negligible. On the contrary, the disc is still forming stars at the present time and therefore it contributes to the observed masses of the two radionuclides. In our model, the SFR is, in fact, active in the thin-disc now and in agreement with the present time observed values (see later).

	\begin{figure}
		\centering
		\includegraphics[scale=0.5]{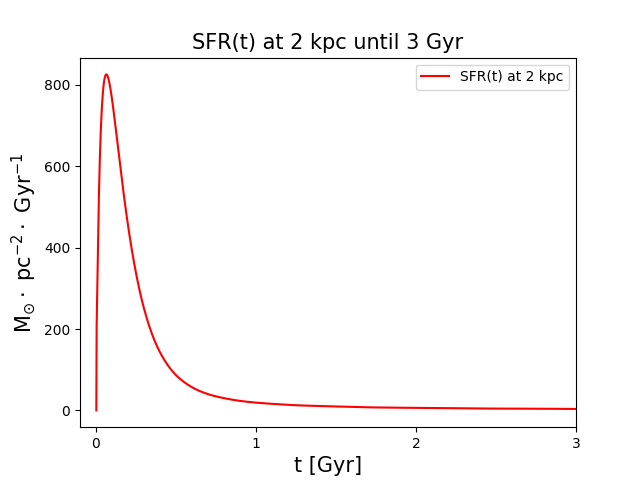}
		\caption{SFR(t) at $2$ kpc as a function of time in the first $3$ Gyr since the beginning of star formation. The prescriptions assumed (high star formation efficiency, short time scale and single infall episode) produce a SFR extremely high at the beginning of the Galaxy which decreases immediately due to the fast gas consumption.}
		\label{fig:sfr2kpczoom}
	\end{figure}

	\subsection{Chemical Evolution Equations}
	Our previous model has been here improved by including chemical evolution equations for radioactive elements, in particular, for a given element $i$ we have the following equation:
	\begin{equation}
	\frac{d(X_{i}M_{gas})}{dt}=-X_{i}(t)\psi(t)+R_{i}(t)+X_{A}A(t)-\overbrace{\lambda_{i}X_{i}(t)M_{gas}}^{\text{Radioactive decay}}
	\end{equation}
	where $X_{i}(t)\psi(t)$ is the rate at which the element $i$ is included in new stars that are forming at the time $t$, $R_{i}(t)$ accounts for the injection rate of the element $i$ in the ISM and $X_{A}A(t)$ represents the infall contribution. The last term, $\lambda_{i}X_{i}(t)M_{gas}$, where $\lambda_{i}$ is the decay constant of the element $i$, accounts for the radioactive decay in the ISM. The decay constant is the inverse of the decay timescale and we have $\tau_{26Al}=1.05$ Myr and $\tau_{60Fe}= 3.75$ Myr (Diehl $2013$).
	\newline
	The radioisotopes  decay also in the stellar interior, and this fact is taken into account in the equations through the production matrix $Q_{mi}(t-\tau_{m})$, involved in the calculation of the contribution by each kind of star. The diagonal terms of such a matrix represent the fraction of elements which are not processed by the star, but that are ejected in the ISM in their original form. For a radioactive nucleus this fraction decreases during the life of the star due to the decay: to consider this phenomenon we added an exponential factor $e^{-\lambda_{i}\tau_{m}}$ to the diagonal terms of the matrix.
	\newline
	The most noteworthy aspect of equation ($10$) is that it is valid both for stable and unstable nuclei. In fact, for stable nuclei $\lambda_{i}=0$, so the decay term in equation ($10$) is zero and the exponential factor added in the production matrix equals $1$.
	
	\subsection{Nucleosynthesis and yields}
	Among the input parameters, the nucleosynthesis plays a major role in this study. To account for the production of $^{26}$Al and $^{60}$Fe we considered the contribution by massive stars (M$\geq 13$M$_{\odot}$), AGB stars (M$\leq 6$M$_{\odot}$), S-AGB stars ($6$ M$_{\odot}<$M$<13$ M$_{\odot}$), SNIa and, only for $^{26}$Al, nova systems.
	\newline
	For massive stars (M$\geq13$M$_{\odot}$), which are the main producers of the two studied nuclides, we tested four different yield sets: Woosley \& Weaver ($1995$) with initial metallicity dependence (WW$95$Zdep), Woosley \& Weaver ($1995$) at the solar metallicity only (WW$95$Z$_{\odot}$), Limongi \& Chieffi ($2006$) (LC$06$), and Limongi \& Chieffi ($2018$) (LC$18$). The set of yields by LC$18$ is also dependent of stellar rotation. The authors provide yields for three different rotational velocities, 0 km s$^{-1}$, $150$ km s$^{-1}$ and $300$ km s$^{-1}$. We implemented this set in our model assuming that all stars with [Fe/H]$\leq -1$ dex are fast rotators whereas those with [Fe/H]$>-1$ dex are non-rotating, as suggested by Romano et al. ($2019$). For $^{60}$Fe, LC$06$ offers two different models depending on the convection criterion adopted, namely Schwarzschild or Ledoux criterion. In this study we tested both of them, for a total of five sets of massive star yields for $^{60}$Fe and only four sets for $^{26}$Al.
	\newline
	For AGB stars (M$\leq6$ M$_{\odot}$) we adopted the yields by Karakas ($2010$) and for S-ABG stars ($6$M$_{\odot}<$M$<13$M$_{\odot}$) those by Doherty et al. ($2014$a,b). We underline that for stars within the mass interval $10$-$13$ M$_{\odot}$ we assumed constant yields because Doherty et al. ($2014$a,b) do not provide values for this mass range.
	\newline
	To account for the contribution by SNIa we assumed the yields by Nomoto, Thielemann \& Yokoi ($1984$).
	\newline
	Finally, we considered three different sets of yields for the nova system production of $^{26}$Al: one by Jos\'{e} \& Hernanz ($1998$) (Model B) and two by Jos\'{e} \& Hernanz ($2007$) (Model A and Model C). In particular, in Jos\'{e} \& Hernanz ($1998$) seven models for CO novae and seven models for ONe novae are listed. We selected the best one following the suggestion by Romano \& Matteucci ($2003$). We assumed that the $30\%$ of the nova systems are ONe novae and the remaining $70\%$ are CO novae, and the best model for each kind of novae is the average among the seven available:
	\begin{equation}
	\langle X_{^{26}Al}\rangle = 0.7\langle X_{^{26}Al}\rangle_{CO}+0.3\langle X_{^{26}Al}\rangle_{ONe}
	\end{equation}
	In addition, we also computed a fourth model without nova production.
	Being the nova nucleosynthesis independent of that by massive stars, for $^{26}$Al we tested the four massive star yields all combined with the four nova sets of yields, for a total of sixteen yield sets tested. The values of the mass of $^{26}$Al produced in a nova outburst are listed in Table 1. These values are then multiplied by the expected number of outbursts during the lifetime of a nova system, that we assume to be $10^{4}$ (refs.).
	\newline
	For $^{60}$Fe, as anticipated, no nova production is assumed, so we tested only the five massive star sets of yields.
	\newline
	The yield sets are shown in Table $1$. In the first column, the models are listed, in the second column is reported the reference and the third column contains details of different models belonging to the same reference.
	
	\begin{table*}
		\caption{Models for different massive star yields. The first column reports the model identification, the second column contains the reference and the third column is used for the details of the models, to distinguish those taken from the same reference. For nova yields (only for $^{26}$Al), in the third column are listed the adopted values. These values are the production during one outburst. We then assume $10^{4}$ outbursts in total during the life of a nova system.}
		\begin{center}
			\begin{tabular}{c|c|c}
				\hline
				\multicolumn{1}{c}{\textbf{Model}} & \multicolumn{1}{|c|}{\textbf{Reference}}& \multicolumn{1}{c}{\textbf{Characteristics}}\\
				\multicolumn{3}{>{\columncolor[gray]{.8}}c}{\textbf{Massive stars}} \\
				\multicolumn{1}{c}{Model $1$}&\multicolumn{1}{|c|}{Woosley \& Weaver ($1995$)}&\multicolumn{1}{c}{metallicity dependent (Zdep)}\\
				\hline
				\multicolumn{1}{c}{Model $2$}&\multicolumn{1}{|c|}{Woosley \& Weaver ($1995$)}&\multicolumn{1}{c}{solar metallicity (Z$_{\odot}$)}\\
				\hline
				\multicolumn{1}{c}{Model $3$}&\multicolumn{1}{|c|}{Limongi \& Chieffi ($2006$)}&\multicolumn{1}{c}{Schwarzschild criterion}\\
				\hline
				\multicolumn{1}{c}{Model $4$}&\multicolumn{1}{|c|}{Limongi \& Chieffi ($2006$)}&\multicolumn{1}{c}{Ledoux criterion}\\
				\hline
				\multicolumn{1}{c}{Model $5$}&\multicolumn{1}{|c|}{Limongi \& Chieffi ($2018$)}&\multicolumn{1}{c}{}\\
				\multicolumn{3}{>{\columncolor[gray]{.8}}c}{\textbf{Nova systems (only for $^{26}$Al)}} \\
				\multicolumn{1}{c}{Model A}&\multicolumn{1}{|c|}{Jos\'{e} \& Hernanz ($2007$)}&\multicolumn{1}{c}{low production (1.38E-4 M$_{\odot}$)}\\
				\hline
				\multicolumn{1}{c}{Model B}&\multicolumn{1}{|c|}{Jos\'{e} \& Hernanz ($1998$)}&\multicolumn{1}{c}{intermediate production (3.82E-4 M$_{\odot}$)}\\
				\hline
				\multicolumn{1}{c}{Model C}&\multicolumn{1}{|c|}{Jos\'{e} \& Hernanz ($2007$)}&\multicolumn{1}{c}{high production (7.69E-4 M$_{\odot}$)}\\
				\hline
			\end{tabular}
		\end{center}
	\end{table*}
	
	%For $^{26}$Al, the combination of massive star yields with nova system yields is labeled as described in Table $2$. The first column refers to the final label, the second to the massive star yields, and in the third column the nova yields are listed. We remind that the models labeled as \textquotedblleft Model $3$\textquotedblright $\,$ and \textquotedblleft Model $4$\textquotedblright $\,$ are different only for $^{60}$Fe due to the different convection criteria adopted, whereas for $^{26}$Al they produce the same results. In Table $2$, when combining the massive star production with the nova production of $^{26}$Al, \textquotedblleft Model $4$\textquotedblright $\,$ has been neglected because no differences exist with \textquotedblleft Model $3$\textquotedblright.
	
	\begin{table*}
		\caption{Final models, combining massive stars and nova yields, for $^{26}$Al. The first column reports the label given to the models with combined yields. The second and third columns contains the prescriptions adopted for massive star yields and nova yields, respectively. \textquotedblleft Model $4$\textquotedblright $\,$ is missing because for $^{26}$Al it is equivalent to \textquotedblleft Model $3$\textquotedblright. These two models are different only for $^{60}$Fe.}
		\begin{center}
			\begin{tabular}{c|c|c}
				\hline
				\multicolumn{3}{>{\columncolor[gray]{.8}}c}{\textbf{Combined models (only for $^{26}$Al)}} \\
				\multicolumn{1}{c|}{\textbf{Final label}}&\multicolumn{1}{c}{\textbf{Massive stars}} & \multicolumn{1}{|c}{\textbf{Nova systems}}\\
				\hline
				\multicolumn{1}{c|}{Model 1}&\multicolumn{1}{c}{Model 1}&\multicolumn{1}{|c}{no production}\\
				\hline
				\multicolumn{1}{c|}{Model 1A}&\multicolumn{1}{c}{Model 1}&\multicolumn{1}{|c}{Model A}\\
				\hline
				\multicolumn{1}{c|}{Model 1B}&\multicolumn{1}{c}{Model 1}&\multicolumn{1}{|c}{Model B}\\
				\hline
				\multicolumn{1}{c|}{Model 1C}&\multicolumn{1}{c}{Model 1}&\multicolumn{1}{|c}{Model C}\\
				\multicolumn{3}{>{\columncolor[gray]{.8}}c}{}\\
				\multicolumn{1}{c|}{Model 2}&\multicolumn{1}{c}{Model 2}&\multicolumn{1}{|c}{no production}\\
				\hline
				\multicolumn{1}{c|}{Model 2A}&\multicolumn{1}{c}{Model 2}&\multicolumn{1}{|c}{Model A}\\
				\hline
				\multicolumn{1}{c|}{Model 2B}&\multicolumn{1}{c}{Model 2}&\multicolumn{1}{|c}{Model B}\\
				\hline
				\multicolumn{1}{c|}{Model 2C}&\multicolumn{1}{c}{Model 2}&\multicolumn{1}{|c}{Model C}\\
				\multicolumn{3}{>{\columncolor[gray]{.8}}c}{}\\
				\multicolumn{1}{c|}{Model 3}&\multicolumn{1}{c}{Model 3}&\multicolumn{1}{|c}{no production}\\
				\hline
				\multicolumn{1}{c|}{Model 3A}&\multicolumn{1}{c}{Model 3}&\multicolumn{1}{|c}{Model A}\\
				\hline
				\multicolumn{1}{c|}{Model 3B}&\multicolumn{1}{c}{Model 3}&\multicolumn{1}{|c}{Model B}\\
				\hline
				\multicolumn{1}{c|}{Model 3C}&\multicolumn{1}{c}{Model 3}&\multicolumn{1}{|c}{Model C}\\
				\multicolumn{3}{>{\columncolor[gray]{.8}}c}{}\\
				\multicolumn{1}{c|}{Model 5}&\multicolumn{1}{c}{Model 5}&\multicolumn{1}{|c}{no production}\\
				\hline
				\multicolumn{1}{c|}{Model 5A}&\multicolumn{1}{c}{Model 5}&\multicolumn{1}{|c}{Model A}\\
				\hline
				\multicolumn{1}{c|}{Model 5B}&\multicolumn{1}{c}{Model 5}&\multicolumn{1}{|c}{Model B}\\
				\hline
				\multicolumn{1}{c|}{Model 5C}&\multicolumn{1}{c}{Model 5}&\multicolumn{1}{|c}{Model C}\\
				\hline
			\end{tabular}
		\end{center}
	\end{table*}
	
	For $^{26}$Al, the combination of massive star yields with nova system yields is labeled as described in Table $2$. The first column refers to the final label, the second to the massive star yields, and in the third column the nova yields are listed. We remind that the models labeled as \textquotedblleft Model $3$\textquotedblright $\,$ and \textquotedblleft Model $4$\textquotedblright $\,$ are different only for $^{60}$Fe due to the different convection criteria adopted, whereas for $^{26}$Al they produce the same results. In Table $2$, when combining the massive star production with the nova production of $^{26}$Al, \textquotedblleft Model $4$\textquotedblright $\,$ has been neglected because no differences exist with \textquotedblleft Model $3$\textquotedblright.
	
	\section{Results}
	Here we show the results obtained with the models described above. For each model we computed the integrated mass and the integrated injection rates of $^{26}$Al and $^{60}$Fe as a function of Galactocentric radius, and, where possible, we compared these theoretical results with the observations.
	
	\subsection{$^{26}$Al mass}
	For $^{26}$Al the surveys COMPTEL and INTEGRAL observed a mass in the range $1.8-3.6$ M$_{\odot}$ within $5$ kpc from the Galactic centre, so we considered each model within this interval as an acceptable one. 
	In Table $3$ we listed the masses of $^{26}$Al computed within three significant Galactocentric radii: $5$ kpc (the observations scale radius), $8$ kpc (the solar neighborhood) and $18$ kpc (the outer border of the Galaxy). We compared the value at $5$ kpc with the observations, in order to highlight the best models.
	From Table $3$ we notice that the models without nova production, those with low nova production (Model A) and with moderate nova production (Model B) produce too low  mass of $^{26}$Al, independent of the massive star yields.
	On the contrary Model C (high nova contribution) always agrees with the observations.
	The four compatible models are Model $1$C, Model $2$C, Model $3$C and Model $5$C and are all shown in Figure $3$, together with the observed interval (thick yellow line) and three non-fitting models, Model $2$, Model $2$A and Model $2$B, chosen as representative examples. In addition, the red dot at $2$ M$_{\odot}$ represents the most plausible observation of $^{26}$Al at $5$ kpc according to Diehl et al. ($2010$) and Diehl ($2016$). By means of Figure $3$ we were able to select the best model: among the four compatible ones, Model $1$C (dash-dotted red line) is the best, with $2.12$ M$_{\odot}$ of $^{26}$Al produced, which is the closest value to the observations. These results suggest that the nova contribution to $^{26}$Al cannot be avoided in order to reproduce the observations.
	
	\begin{figure}
		\includegraphics[scale=0.5]{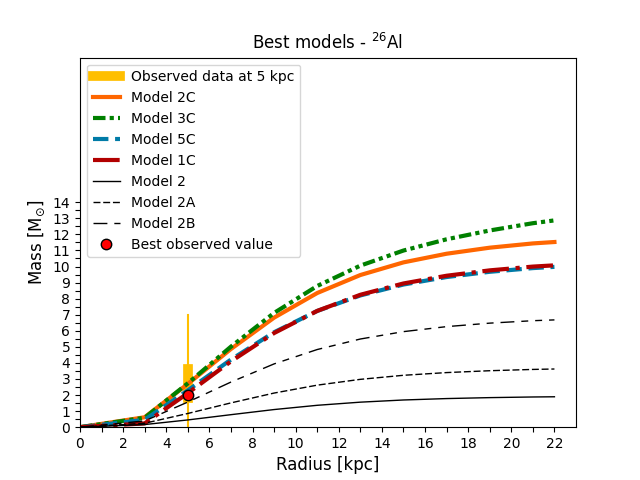}
		\label{fig:massaalbestmod}
		\caption{Integrated mass of $^{26}$Al as a function of Galactocentric distance for seven odels among the computed ones. Model $1$C (dash-dotted red line), Model $2$C (solid orange line), Model $3$C (dash-dot-dotted green line ) and Model $5$C (dashed blue line) are the only four compatible models we obtained. The three black lines are Model $2$, Model $2$A and Model $2$B, plotted as examples of non compatible models (they were chosen as the highest model among those which do not fit the observations). We stress that Model $2$ witout novae is a factor 5 lower than the compatible model. The yellow band represents the observed values and the red dot is the best observation. Although $1$C, $2$C, $3$C and $5$C all offer a mass value within $1.8-3.6$ M$_{\odot}$ at $5$ kpc, Model $1$C is the closest to the best observation, $\sim 2$ M$_{\odot}$.}
	\end{figure}

	\begin{table*}
		\caption{Integrated mass of $^{26}$Al within three different Galactocentric radii: $5$ kpc, $8$ kpc and $18$ kpc. We compared the mass within $5$ kpc with that observed by INTEGRAL, which lies in the range $1.8-3.6$ M$_{\odot}$. Only four models are in agreement with the observed interval: Model $1$C, Model $2$C, Model $3$C and Model $5$C.}
		\begin{centering}
			\begin{tabular}{c|c|c|c|c|}
				\hline
				\multicolumn{1}{c}{} & \multicolumn{4}{|c}{} \\
				\multicolumn{1}{c}{} & \multicolumn{4}{|c}{\textbf{\large $^{26}$Al results}}\\
				\multicolumn{1}{c}{} & \multicolumn{4}{|c}{} \\
				\hline
				\multicolumn{1}{l}{\textbf{Radius [kpc]}} & \multicolumn{1}{|c}{\textbf{Model 1 [$M_{\odot}$]}} & \multicolumn{1}{|c}{\textbf{Model 1A [$M_{\odot}$]}}& \multicolumn{1}{|c}{\textbf{Model 1B [$M_{\odot}$]}}&\multicolumn{1}{|c}{\textbf{Model 1C [$M_{\odot}$]}}\\
				\hline
				\multicolumn{1}{l}{\textbf{5}}&\multicolumn{1}{|c}{0.10}&\multicolumn{1}{|c}{0.50}&\multicolumn{1}{|c}{1.22}&\multicolumn{1}{|c}{2.12}\\
				\hline
				\multicolumn{1}{l}{\textbf{8}}&\multicolumn{1}{|c}{0.36}&\multicolumn{1}{|c}{1.38}&\multicolumn{1}{|c}{3.20}&\multicolumn{1}{|c}{5.86}\\
				\hline
				\multicolumn{1}{l}{\textbf{18}}&\multicolumn{1}{|c}{0.65}&\multicolumn{1}{|c}{2.32}&\multicolumn{1}{|c}{5.29}&\multicolumn{1}{|c}{9.76}\\
				\hline
				\multicolumn{5}{>{\columncolor[gray]{1.0}}c}{} \\
				\hline
				\multicolumn{1}{l}{\textbf{Radius [kpc]}}&\multicolumn{1}{|c}{\textbf{Model 2 [$M_{\odot}$]}}&\multicolumn{1}{|c}{\textbf{Model 2A [$M_{\odot}$]}}&\multicolumn{1}{|c}{\textbf{Model 2B [$M_{\odot}$]}}&\multicolumn{1}{|c}{\textbf{Model 2C[$M_{\odot}$]}}\\
				\hline
				\multicolumn{1}{l}{\textbf{5}}&\multicolumn{1}{|c}{0.45}&\multicolumn{1}{|c}{0.85}&\multicolumn{1}{|c}{1.56}&\multicolumn{1}{|c}{2.69}\\
				\hline
				\multicolumn{1}{l}{\textbf{8}}&\multicolumn{1}{|c}{1.10}&\multicolumn{1}{|c}{2.12}&\multicolumn{1}{|c}{3.94}&\multicolumn{1}{|c}{6.82}\\
				\hline
				\multicolumn{1}{l}{\textbf{18}}&\multicolumn{1}{|c}{1.84}&\multicolumn{1}{|c}{3.51}&\multicolumn{1}{|c}{6.48}&\multicolumn{1}{|c}{11.17}\\
				\hline
				\multicolumn{5}{>{\columncolor[gray]{1.0}}c}{}\\
				\hline
				\multicolumn{1}{l}{\textbf{Radius [kpc]}}&\multicolumn{1}{|c}{\textbf{Model 3 [$M_{\odot}$]}}&\multicolumn{1}{|c}{\textbf{Model 3A [$M_{\odot}$]}}&\multicolumn{1}{|c}{\textbf{Model 3B [$M_{\odot}$]}}&\multicolumn{1}{|c}{\textbf{Model 3C [$M_{\odot}$]}}\\
				\hline
				\multicolumn{1}{l}{\textbf{5}}&\multicolumn{1}{|c}{0.38}&\multicolumn{1}{|c}{0.80}&\multicolumn{1}{|c}{1.55}&\multicolumn{1}{|c}{2.73}\\
				\hline
				\multicolumn{1}{l}{\textbf{8}}&\multicolumn{1}{|c}{0.97}&\multicolumn{1}{|c}{2.07}&\multicolumn{1}{|c}{4.02}&\multicolumn{1}{|c}{7.11}\\
				\hline
				\multicolumn{1}{l}{\textbf{18}}&\multicolumn{1}{|c}{1.72}&\multicolumn{1}{|c}{3.60}&\multicolumn{1}{|c}{6.95}&\multicolumn{1}{|c}{12.24}\\
				\hline
				\multicolumn{5}{>{\columncolor[gray]{1.0}}c}{}\\
				\hline
				\multicolumn{1}{l}{\textbf{Radius [kpc]}}&\multicolumn{1}{|c}{\textbf{Model 5 [$M_{\odot}$]}}&\multicolumn{1}{|c}{\textbf{Model 5A [$M_{\odot}$]}}&\multicolumn{1}{|c}{\textbf{Model 5B [$M_{\odot}$]}}&\multicolumn{1}{|c}{\textbf{Model 5C [$M_{\odot}$]}}\\
				\hline
				\multicolumn{1}{l}{\textbf{5}}&\multicolumn{1}{|c}{0.05}&\multicolumn{1}{|c}{0.46}&\multicolumn{1}{|c}{1.17}&\multicolumn{1}{|c}{2.29}\\
				\hline
				\multicolumn{1}{l}{\textbf{8}}&\multicolumn{1}{|c}{0.19}&\multicolumn{1}{|c}{1.22}&\multicolumn{1}{|c}{3.03}&\multicolumn{1}{|c}{5.91}\\
				\hline
				\multicolumn{1}{l}{\textbf{18}}&\multicolumn{1}{|c}{0.34}&\multicolumn{1}{|c}{2.01}&\multicolumn{1}{|c}{4.98}&\multicolumn{1}{|c}{9.67}\\
				\hline
			\end{tabular}
		\end{centering}
	\end{table*}
	
	\subsection{$^{60}$Fe mass}
	\begin{figure}
		\includegraphics[scale=0.5]{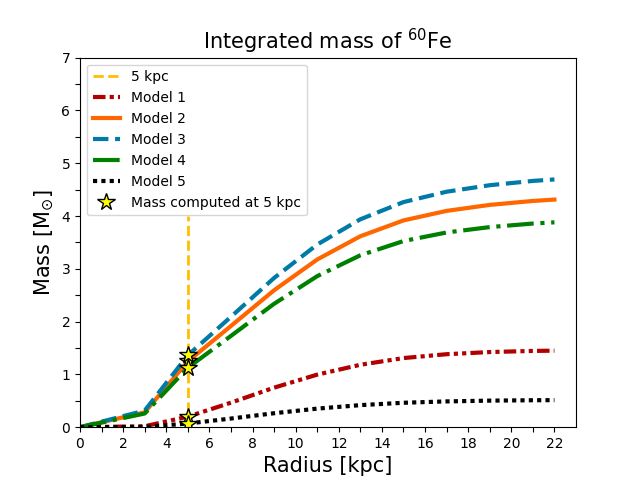}
		\label{fig:massafe}
		\caption{Integrated mass of $^{60}$Fe as a function of Galactocentric radius obtained with the five models tested. It is evident that Model $1$ (dash-dot-dotted red line) and Model $5$ (dotted black line), which are the two metallicity dependent ones, produce the lowest $^{60}$Fe masses, whereas Model $2$, Model $3$ and Model $4$ produce higher masses of $^{60}$Fe. According to the conversion between flux ratio and abundance ratio, $^{60}$Fe should lie in the range $0.9-1.8$ M$_{\odot}$. Our best model is Model $3$ with $1.05$ M$_{\odot}$ produced within $5$ kpc from the Galactic centre.}
	\end{figure}
	
	Also for $^{60}$Fe we computed the integrated mass within $5$ kpc, $8$ kpc and $18$ kpc, that are listed in Table $4$. Nevertheless, we did not perform an observational comparison, due to the absence of a $^{60}$Fe mass estimate. The computed values can work as predictions and constraints for future observations. In Figure $4$ we show the integrated masses of $^{60}$Fe as functions of the Galactocentric distance for all five models, with the yellow stars representing the predicted mass of $^{60}$Fe at $5$ kpc for each model. Model $1$ (dash-dot-dotted red line) and Model $5$ (dotted black line) are much lower than the other three, due probably to their metallicity dependence of the yields. Model $2$, Model $3$ and Model $4$ differ for the input prescriptions but the results they offer are similar.
	\newline
	To identify the best model we considered the flux ratio $^{60}$Fe/$^{26}$Al observed by INTEGRAL. INTEGRAL measured a ratio in the range $0.2-0.4 $ which correspons to a mass ratio that lies in the range $0.46-0.92$.  Therefore, by assuming that the best $^{26}$Al mass observation is $2$ M$_{\odot}$, $^{60}$Fe mass should lie in the range $0.9-1.8$ M$_{\odot}$. A ratio lower than unity means that the mass of $^{60}$Fe is lower than that of $^{26}$Al. We stress that Model $3$ is extremely noteworthy from this point of view: it is the only compatible model among those tested, since it produces $1.05$ M$_{\odot}$ of $^{60}$Fe within $5$ kpc form the Galactic centre. Given that we can consider Model $3$ (Limongi \& Chieffi $2006$ with Schwarzschild production criterion) as the best model for $^{60}$Fe.
	
	\begin{table*}
		\caption{Integrated mass of $^{60}$Fe within $5$ kpc, $8$ kpc and $18$ kpc for the five massive star models tested. It is not possible to perform a comparison with the $\gamma$ observations because no mass estimate are available for $^{60}$Fe. The results listed here are a prediction of the $^{60}$Fe mass at different Galactocentric distances, and can be used as constraints for future $\gamma$-ray surveys.}
		\begin{centering}
			\begin{tabular}{c|c|c|c|c|c}
				\hline
				\multicolumn{1}{c|}{} & \multicolumn{5}{c}{}\\
				\multicolumn{1}{c|}{} & \multicolumn{5}{c}{\textbf{\large $^{60}$Fe results}}\\
				\multicolumn{1}{c|}{} & \multicolumn{5}{c}{}\\
				\hline
				\multicolumn{1}{l|}{\textbf{Radius [kpc]}}&\multicolumn{1}{c|}{\textbf{Model 1 [$M_{\odot}$]}}&\multicolumn{1}{c|}{\textbf{Model 2 [$M_{\odot}$]}}&\multicolumn{1}{c|}{\textbf{Model 3 [$M_{\odot}$]}}&\multicolumn{1}{c|}{\textbf{Model 4 [$M_{\odot}$]}}&\multicolumn{1}{c}{\textbf{Model 5 [$M_{\odot}$]}}\\
				\hline
				\multicolumn{1}{l|}{\textbf{5}}&\multicolumn{1}{c|}{0.20}&\multicolumn{1}{c|}{0.87}&\multicolumn{1}{c|}{1.05}&\multicolumn{1}{c|}{0.82}&\multicolumn{1}{c}{0.07}\\
				\hline
				\multicolumn{1}{l|}{\textbf{8}}&\multicolumn{1}{c|}{0.71}&\multicolumn{1}{c|}{2.17}&\multicolumn{1}{c|}{2.59}&\multicolumn{1}{c|}{2.09}&\multicolumn{1}{c}{0.26}\\
				\hline
				\multicolumn{1}{l|}{\textbf{18}}&\multicolumn{1}{c|}{1.28}&\multicolumn{1}{c|}{3.64}&\multicolumn{1}{c|}{4.52}&\multicolumn{1}{c|}{3.69}&\multicolumn{1}{c}{0.45}\\
				\hline
			\end{tabular}
		\end{centering}
	\end{table*}
	
	\subsection{Injection rates}

	By means of Model $2$, we also computed the present time integrated injection rates (measured in M$_{\odot}$ pc$^{-1}$ Gyr$^{-1}$), for both $^{26}$Al and $^{60}$Fe: they represent the present rate at which these elements are injected into the ISM and are integrated assuming that they are constant in each Galactocentric ring.
	In Figure $5$, we show the comparison between the injection rates computed here and those of Timmes et al. ($1995$).  
	The solid red line and the dashed blue line represent the integrated injection rates we obtained for $^{26}$Al and $^{60}$Fe, respectively. In Model 2, we assumed the same yields and the same IMF (Salpeter ($1955$) normalized from $0.08$ M$_{\odot}$ to $40$ M$_{\odot}$), as in Timmes et al. ($1995$), in order to compare with their results. In Figure 5, are reported also the results of Timmes et al.($1995$): the dotted red line and dash-dotted blue line are the integrated injection rates for $^{26}$Al and $^{60}$Fe, respectively. 
	\newline
	The plot shows clearly that in both cases the injection rate of $^{26}$Al is higher than that of $^{60}$Fe, but the trends are different in the two papers. 
	The differences between our results and the previous ones, can be ascribed to different assumptions in the models adopted for the chemical evolution of the Milky Way. In particular, in Timmes et al. ($1995$), they considered only one infall episode with a time scale constant with Galactocentric distance, and they assumed that the Galactic bulge evolves as the innermost thin-disc.
	These two hypothesis are different from those used in this present study, where we assume an inside-out formation of the thin disc and a separate model for the bulge. These differences are probably responsible for the disagreement among the plots.
	%The reason lies in the fact that the $\tau_{D}(R)$ value assumed by Timmes et al. ($1995$) is constant, whereas we assumed that $\tau_{D}(R)$ increases with R, as explained in Section $2.1$. Moreover, in Timmes et al. ($1995$) the Galactic bulge evolves as the innermost disc and therefore they predict active star formation at the present time. The most important consequence of the facts just described, is the location of the maximum in the plot injection rate vs. Galactocentric distance which in Timmes et al. ($1995$) is in the bulge. Actually, it is not very likely that the highest contribution comes from the bulge because of the negligible SFR at the present time: being $^{26}$Al and $^{60}$Fe tracers of active star formation because produced mainly by massive stars, the highest contribution should originate from the innermost disc, where the SFR is the highest (\textbf{REFERENCES}).
	In particular, our results show a maximum located at around $5-7$ kpc which corresponds roughly to the maximum in the present time SFR, exactly as it was expected. In the bulge, where the current SFR is around $\sim0.2$ M$_{\odot}$ pc$^{-2}$ Gyr$^{-1}$, the injection rates are negligible. In the same way, at larger radii the injection rates decrease due to the assumed exponential Galactic disc and inside-out formation which lower the SFR, as shown in Figure $6$, where we report the predicted present time SFR gradient along the thin disc. 
	
	\begin{figure}
		\includegraphics[scale=0.5]{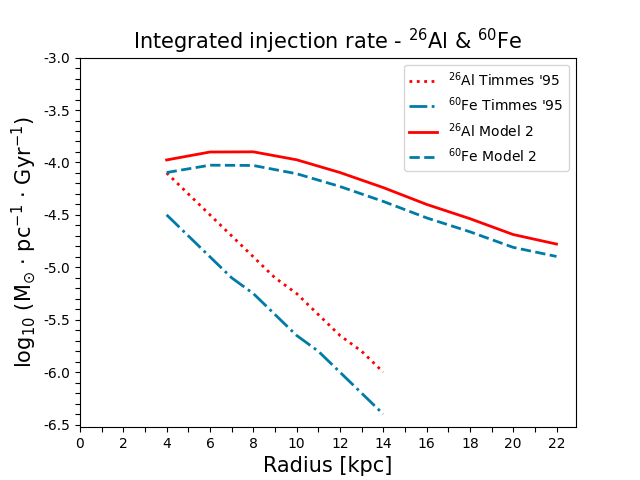}
		\label{fig:wijalfe}
		\caption{Present time integrated injection rates computed with Model $2$ for $^{26}$Al (solid red line) and $^{60}$Fe (dashed blue line), compared with those by Timmes et al. ($1995$) for $^{26}$Al (dotted red line) and for $^{60}$Fe (dash-dotted blue line) only for the disc (R$\geq 4$ kpc). 
			\newline
			\textit{Timmes et al. ($1995$)}: the results are computed assuming that the time scale of the second infall $\tau_{D}$ is constant (no \textquotedblleft inside-out\textquotedblright scenario assumed) and that the IMF is that by Salpeter ($1955$).
			\newline
			\textit{This study}: in this case we assumed the \textquotedblleft inside-out\textquotedblright scenario for the disc and a Salpeter ($1955$) IMF. The results we get show that the highest contribution comes from the $6-8$ kpc region. The differences between the two studies do not lay in the IMF but in other model prescriptions such as the inside-out scenario.} %In both cases the rate of $^{26}$Al is higher, but the trends are different due to the law assumed for the second infall time scale $\tau_{D}(R)$. $^{26}$Al and $^{60}$Fe are tracers of active star formation, so the highest contribution is expected from the disc. In addition Timmes et al. ($1995$) assumed that the bulge evolves as the innermost disc, predicting a high present time SFR(t). Moreover, they assumed a Miller \& Scalo ($1979$) IMF normalized from $0.08$ M$_{\odot}$ to $40$ M$_{\odot}$ both for the bulge and the disc. In this study, as explained in Section $2.1$, on the contrary, we assumed a three-slopes Kroupa ($1993$) IMF for the disc and a Salpeter ($1995$) for the bulge from $0.1$ M$_{\odot}$ to $100$ M$_{\odot}$. Our prescriptions predict a low current SFR(t) ($\sim 0.2$ M$_{\odot}$ pc$^{-2}$ Gyr$^{-1}$), in agreement with the expectation. Regarding the injection rates, the maximum of both is located around $4-5$ kpc, whereas those by Timmes et al. ($1995$) do not (the maxima are in the bulge).}
	\end{figure}
	
	\begin{figure}
		\centering
		\includegraphics[scale=0.5]{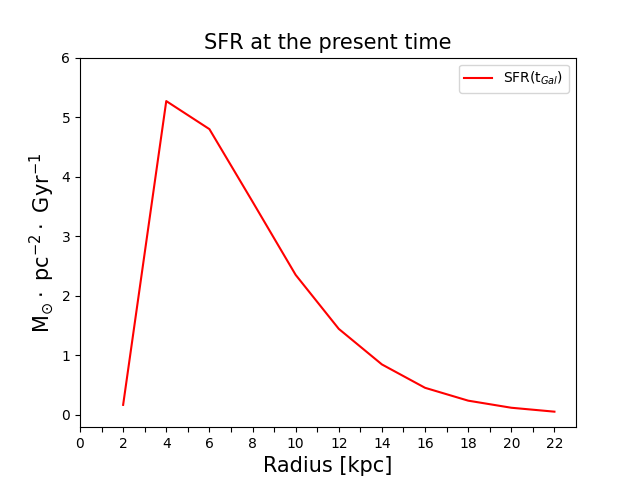}
		\caption{SFR at the present time as a function of Galactocentric distance assumed in the model. The maximum is located at $4$ kpc from the Galactic centre, whereas in the bulge and in the outer Galaxy it decreases.}
		\label{fig:sfrtgal}
	\end{figure}
	
	\section{Discussion on the model parameters}

	It is worth noting that in this paper we have only varied the stellar yields for the two radionuclides, $^{26}$Al and $^{60}$Fe. The reason is that we adopted state-of-the-art chemical evolution models containing SFR, IMF and infall laws already tested in previous papers. In particular, the model for the bulge assumes a very intense SFR leading to a quick gas consumption, a hypothesis that is required to reproduce the stellar metallicity distribution as well as the [X/Fe] vs [Fe/H] diagrams for a large number of chemical elements, such as $\alpha$-elements (Cescutti \& Matteucci $2011$, Matteucci et al. $2019$), as well as s- and r-process elements (Grisoni et al. 2020). The model for thick- and thin discs with the two-infall framework has also been tested in several previous papers (e.g. Grisoni et al. $2017$, Spitoni et al. $2019$, Palla et al. $2020$). This model reproduces the  [X/Fe] vs. [Fe/H] diagrams for the same elements as those quoted before, plus the Solar System abundances, namely the abundances of the ISM at the time of formation of the Sun (Asplund et al. $2009$; Lodders, $2010$), as well as several present time observational constraints (see Table $5$).
	%In this work we are concentrated in reproducing the present time masses of two radioactive nuclides and derive constraints on the stars which produce them. Besides massive stars, we have tested the roles of AGB stars, Type Ia SNe and, for the first time, novae. To do that , we have started from state-of-the-art models already tested on other observational constraints.
	In Table $5$, we show a comparison between some observational constraints and the predictions of our model for the solar vicinity (SFR, surface gas density, surface stellar density, total surface mass density) and for the whole disc (gas infall rate, SN rates, nova rate). The first column lists the physical quantity considered, the second shows the value predicted by our model, the third contains the observed value and in the fourth column we list the references for the observed values.
	%Moreover, as additional constraint, we predict a Fe solar abundance (the abundance in the ISM at the time of foprmation of the Solar System) of $7.47$ dex, where the photospheric abundance of the Sun by Asplund et al. ($2009$) is $7.50\pm 0.04$ dex and that by Lodders ($2010$) is $7.45\pm 0.08$ dex.
	We can see that, the values we obtain for the physical quantities in Table $5$, are all in reasonable agreement with the observations, and this confirms the initial choice of the parameters.

	\begin{table*}
		\caption{Values of the most important observational constraint predicted by our model. Some quantities refer to the solar vicinity(surface density of gas, alive stars, total surface mass density and star formation rate) and others are averaged over the whole disc (SNIa rate, SNII rate, nova rate and gas infall rate). The first column specifies the physical quantity considered, the second contains the prediction given by our model, the third contains the observed present day value and the last one lists the reference of the observation.}
		\begin{center}
			
			\begin{tabular}{l|c|c|c}
				\hline
				Observable & Predicted value & Observed value & References\\
				%		& & & \\
				\hline
				Surface density of gas & $8.5$ $M_{\odot}$ pc$^{-2}$  & $13\pm3$ $M_{\odot}$ pc$^{-2}$ & Kulkarni \& Heiles ($1987$) \\
				&  & $7$ $M_{\odot}$ pc$^{-2}$ & Dickey ($1993$) \\
				%		& & & \\
				\hline
				Stars (alive) & $\sim 36$ $M_{\odot}$ pc$^{-2}$ & $35\pm5$ $M_{\odot}$ pc$^{-2}$ & Gilmore et al. ($1995$) \\
				& & 33.4 $\pm$ 3 M$_{\odot}$ pc$^{-2}$ & McKee et al. ($2015$) \\
				\hline
				total (discs) & $\sim$54 $M_{\odot}$ pc$^{-2}$ & 48$\pm$ 9 $M_{\odot}$ pc$^{-2}$ & Kuijken \& Gilmore (1991)\\
				%		& & 84$^{+30}$$_{-25}$ $M_{\odot}$ pc$^{-2}$& Bahcall et al. $1992$ \\
				& & 52$\pm$ 13 $M_{\odot}$ pc$^{-2}$& Flynn \& Fuchs (1994) \\
				& & 50-60 $M_{\odot}$ pc$^{-2}$& Cr\'{e}z\'{e} et al. (1998)\\
				%		& & & \\
				\hline
				Star formation rate & $\sim$ 3 $M_{\odot}$ pc$^{-2}$ Gyr$^{-1}$  & 2-10 $M_{\odot}$ pc$^{-2}$ Gyr$^{-1}$ & G\"{u}sten \& Mezger (1982) \\
				& & $\sim$ 5 M$_{\odot}$ pc$^{-2}$ Gyr$^{-1}$ & Prantzos et al. ($2018$) \\
				%		& & & \\
				\hline
				SNeIa rate & 0.43 century$^{-1}$ & 0.3$\pm$0.2 century$^{-1}$ & Cappellaro \& Turatto (1996) \\
				% &  &  & \\
				\hline
				SNeII rate & 1.93 century$^{-1}$ & 1.2$\pm$0.8 century$^{-1}$ & Cappellaro \& Turatto (1996) \\
				% & & &  \\
				\hline
				Novae rate & $\sim$ 43 yr$^{-1}$ & 27-81 yr$^{-1}$ & Shafter (2017) \\
				& & 22-49 yr$^{-1}$ & Darnley et al. (2006) \\
				& & 35$\pm$11 yr$^{-1}$ & Shafter (1997) \\
				& & 20-40 yr$^{-1}$ & Della Valle \& Izzo (2020) \\
				\hline
				Infall rate & $\sim$ 1.5 $M_{\odot}$ pc$^{-2}$ Gyr$^{-1}$ & 0.3-1.5 $M_{\odot}$ pc$^{-2}$ Gyr$^{-1}$& Portinari et al. (1998)\\
				%		& & & \\
				\hline
			\end{tabular}
		\end{center}
	\end{table*}

	\section{Discussion on the yields}
	
	The differences between the yields of Woosley \& Weaver (1995) and Limongi \& Chieffi (2006) that we haave adopted in this paper, does not rely on the nucleosynthesis assumptions but rather on the dependence upon metallicity. As shown in Figure 3, the highest production of $^{26}$Al is obtained when solar metallicity is assumed (Model 2C and Model 3C), whereas a lower production comes from models metallicity dependent (Model 1C). Also yields by Limongi \& Chieffi (2018) with rotation included (Model 5C) offer a lower production of $^{26}$Al, due both to the dependence on metallicity and the effects of rotation.
		
	Other sets of yields for $^{26}$Al and $^{60}$Fe, both for massive stars and for novae, have been provided in the last years. Here, we did not test them but we can perform some qualitative comparisons to understand their main differences relative to those adopted here. For massive stars, Woosley \& Heger (2007) predict a mass ratio $^{60}$Fe/$^{26}$Al almost equal to that by Woosley \& Weaver (1995), therefore we expect similar results. On the other hand, Sukhbold et al. (2016) predict the mass ratio $^{60}$Fe/$^{26}$Al to be lower than that of Woosley \& Weaver (1995) by a factor of 2. As a consequence, we expect that this set of yields would produce proportionately around half of the $^{60}$Fe we produce with Woosley \& Weaver (1995). Regarding the yields by novae, Starrfield et al. (2016) predict a higher ejecta of $^{26}$Al with respect to Jos\`{e} \& Hernanz (1998) (our Model B). Obviously, from this set we expect to have a higher contribution than Model A and Model B, thus reinforcing our conclusion about the importance of $^{26}$Al production by novae.

	\section{Conclusions}
	
	In this study we presented the chemical evolution of two unstable nuclei, $^{26}$Al and $^{60}$Fe, throughout the Milky Way, including the radioactive decay in the chemical evolution equations of models already tested for the thick- and thin- discs and bulge. To account for the production of $^{26}$Al and $^{60}$Fe we considered yields from massive stars, AGB stars, S-AGB stars, SNIa, and for the first time, only for $^{26}$Al, also the contribution from novae. 
	\newline
	For $^{26}$Al we tested sixteen sets of yields, combining four different massive star yields with four different nova yields. For massive stars we used two yield sets by Woosley \& Weaver ($1995$) (metallicity dependent and at solar metallicity), one by Limongi \& Chieffi ($2006$) and one by Limongi \& Chieffi ($2018$). To account for nova production, we tested two models by Jos\'{e} \& Hernanz ($2007$) and one by Jos\'{e} \& Hernanz ($1998$). For each model we computed the mass of $^{26}$Al within three different Galactic radii: $5$ kpc (the scale radius of the observations), $8$ kpc (the solar neighborhood) and $18$ kpc (the outer border of the Galaxy). Then, we compared these values with the $\gamma$-ray observations performed by COMPTEL and INTEGRAL suggesting that the observed mass of $^{26}$Al within $5$ kpc lies in the range $1.8-3.6$ M$_{\odot}$.
	Our results can be summarized as follows:
	
	\begin{itemize}
		\item among the models we tested, only four are compatible with the observed $^{26}$Al mass interval: Model $1$C ($2.12$ M$_{\odot}$), Model $2$C ($2.69$ M$_{\odot}$), Model $3$C ($2.73$ M$_{\odot}$) and Model $5$C ($2.29$ M$_{\odot}$). In addition, the observations indicate that the best $^{26}$Al mass value should be $\sim 2$ M$_{\odot}$. We selected as  best models, Model $1$C and Model $5$C: The first assumes metallicity dependent yields for massive stars by Woosley \& Weaver ($1995$) and high production by nova systems, the second adopts yields from rotating massive stars from Limongi \& Chieffi ($2018$) and high production from nova systems;
		\newline
		\item regarding production of $^{26}$Al by novae, we stress that none of the models without nova production, low nova production (models A) or moderate nova production (models B) is compatible with the observations. This means that in order to reproduce $^{26}$Al observations, the nova contribution is necessary;
		\newline
		\item the effects of AGB and S-AGB stars together with SNe Ia on the production of $^{26}$Al at the present time are negligible;
		\newline
		\item for $^{60}$Fe we tested five massive star sets of yields: two by Woosley \& Weaver ($1995$) (metallicity dependent and at solar metallicity only), two by Limongi \& Chieffi ($2006$) (which differ for the convection criterion adopted) and one by Limongi \& Chieffi ($2018$). In this case,  no comparison with data could be performed due to the absence of $^{60}$Fe mass observations. The only available observational constraint is the flux ratio $^{60}$Fe/$^{26}$Al, which ranges in the interval $0.2-0.4$. According to this, $^{60}$Fe mass should be within $0.9-1.8$ M$_{\odot}$. Model $3$ is the only compatible model: we consider it as best model for $^{60}$Fe with $1.05$ M$_{\odot}$ produced (yields by Limongi \& Chieffi $2006$ with Schwarzschild convection criterion);
		\newline
		%	\item in conclusion, the best theoretical $^{26}$Al mass value within $5$ kpc is $2.12$ M$_{\odot}$, produced by Model $1$C. Regarding $^{60}$Fe, all the models tested underproduce this element, but we can consider as best model Model $3$ (yields by Limongi \& Chieffi $2006$) with $1.05$ M$_{\odot}$ produced. Unfortunately, the yields which offer the best $^{26}$Al estimate are not those which offer the best $^{60}$Fe estimate;
		\newline
		\item 
		finally, we also computed the present time injection rates  for both $^{26}$Al and $^{60}$Fe, assuming the yields used in Model $2$, which contains similar prescriptions to those of Timmes et al. ($1995$).  This was done to compare our rates with those of that previous work. In particular, we adopted the same set of yields for massive stars (e.g. Woosley \& Weaver 1995, solar metallicity) and same IMF (Salpeter, $1955$) but a different Galactic model. This comparison shows that the maximum of our injection rate is located around $5-7$ kpc whereas that by Timmes et al. ($1995$) is located at $4$ kpc.  The differences among the results are probably caused by the different model prescriptions, such as the inside-out scenario in the thin-disc versus a constant timescale for infall.
		
	\end{itemize}
	
	\section*{Acknowledgments}
	
		We thank the referee F.-K. Thielemann for his careful reading and valuable suggestions. We thank R. Diehl for enlightening discussions on the observational characteristics of $^{26}$Al and $^{60}$Fe. We also thank G. Cescutti for suggestions on nucleosynthesis, M. Limongi for valuable hints on radioactive nuclei and D. Romano for careful reading and useful advices. E. Spitoni acknowledges funding from the European Union's Horizon 2020 research and innovation program under SPACE-H2020 grant agreement number 101004214 (EXPLORE project).
	
	\section*{Data availability}
	
		The data underlying this article are available in the article and in its online supplementary material.

\end{document}